\documentclass[a4paper,preprintnumbers,amsmath,amssymb,nofootinbib,showpacs,10pt]{revtex4}
\usepackage[margin=2cm]{geometry}
\usepackage{amssymb}
\usepackage{mathrsfs}
\usepackage{graphics}

\def\P{{\cal{P}}}

\def\nn{\nonumber}
\def\be{\begin{equation}}
\def\ee{\end{equation}}
\def\bea{\begin{eqnarray}}
\def\eea{\end{eqnarray}}
\def\la{\langle}
\def\ra{\rangle}

\def\bk{{\bf k}}
\def\bki{\mathbf k_i}
\def\bkj{\mathbf k_j}

\def\bq{\mathbf q}
\def\bqi{\mathbf q_i}
\def\bqone{\mathbf q_1}
\def\bqtwo{\mathbf q_2}

\def\bki{\mathbf k_i}
\def\bkone{\mathbf k_1}
\def\bktwo{\mathbf k_2}
\def\bkthree{\mathbf k_3}
\def\bkfour{\mathbf k_4}

\def\picube{(2\pi)^3}

\def\variancephi{\la\varphi^2\ra}

\newcommand{\sdelta}[1]{\!\delta^{\,3}(\mathbf{#1})}
\newcommand{\sdeltapi}[1]{(2\pi)^3\!\delta^{\,3}(\mathbf{#1})}

\def\intm{\frac{1}{(2\pi)^3}\int d^3}

\def\Nt{\tilde{N}}
\def\x{\mathbf x}
\def\zetab{\zeta}

\def\variancephi{\la\varphi^2\ra}

\begin{document}

\title{Diagrammatic approach to non-Gaussianity from inflation}
\author{Christian T.~Byrnes$^1$, Kazuya Koyama$^1$, Misao Sasaki$^2$ and David Wands$^1$}

\affiliation{$^1$Institute of Cosmology and Gravitation, Mercantile House, University of
Portsmouth, Portsmouth~PO1~2EG, United Kingdom \\}

\affiliation{$^2$Yukawa Institute for Theoretical Physics, Kyoto University, Kyoto
606-8503, Japan\\}


\pacs{98.80.Cq  \hfill YITP-07-30}

\begin{abstract}
We present Feynman type diagrams for calculating the $n$-point function of the primordial
curvature perturbation in terms of scalar field perturbations during inflation. The
diagrams can be used to evaluate the corresponding terms in the $n$-point function at
tree level or any required loop level. Rules are presented for drawing the diagrams and
writing down the corresponding terms in real space and Fourier space.  We show that
vertices can be renormalised to automatically account for diagrams with dressed vertices.
We apply these rules to calculate the primordial power spectrum up to two loops, the
bispectrum including loop corrections, and the trispectrum.
\end{abstract}

\maketitle

\section{Introduction}

There is currently a great deal of interest in the statistical properties of primordial
perturbations from inflation, because measurements of any non-Gaussianity will improve by
about an order of magnitude over the next few years, for example with the Atacama
Cosmology Telescope \cite{ACT} and Planck \cite{planck}. This will provide a key way to
discriminate between the many models of inflation. Although single field models of
slow-roll inflation typically generate a small level of non-Gaussianity
\cite{Acquaviva:2002ud,Maldacena:2002vr}, there may be an observable level generated in
multiple field inflation \cite{Bassett:2005xm}, for example the curvaton scenario
\cite{curvaton}, or some single field models such as Dirac-Born-Infeld inflation (DBI)
\cite{DBI}. Currently observations of the CMB have concentrated on constraining the
3-point function (bispectrum) \cite{Verde:1999ij,Wang:1999vf,Komatsu:2001rj}, but the
4-point function (trispectrum) has also been considered
\cite{Okamoto:2002ik,Bartolo:2005fp,Kogo:2006kh}, and in principle higher order $n$-point
functions may also be observable if they are sufficiently large. In some models the first
signal of non-Gaussianity may come through the trispectrum, for example some special
cases of the curvaton scenario \cite{byrnes3}. Furthermore higher order statistics in
principle carry more information and therefore could distinguish between different
sources of non-Gaussianity, for example in many cases the bispectrum can be paremetrised
by a single non-linearity parameter, $f_{NL}$, while the trispectrum depends on two
wavenumber independent parameters, $\tau_{NL}$ and $g_{NL}$ \cite{byrnes3}.


We will calculate the primordial curvature perturbation on uniform density hypersurfaces, $\zeta$, on large scales employing the $\delta N$ formalism \cite{Starobinsky:1986fx,Sasaki:1995aw,Lyth:2004gb,Lyth:2005fi}. We
use the separate universe approach \cite{Salopek:1990jq,Sasaki:1998ug,Wands:2000dp}.~This
considers each super-Hubble scale patch to be evolving like a separate
Friedmann-Robertson-Walker universe which are locally homogeneous.~By patching these
regions together we can track the evolution of the perturbations on large scales just by
using background quantities which greatly simplifies our calculations. We also need to
know the perturbations of the scalar fields at Hubble-exit, but in many cases these are
extremely close to Gaussian in which case their statistical properties are given purely
in terms of the power spectrum of the fields.

The number of $e$-foldings, $N$, given by
\bea N=\int^{t_{\rm{fin}}}_{t_{\rm{ini}}}H(t) dt\,, \eea
is evaluated from an initial flat hypersurface to a final uniform-density hypersurface.~The perturbation in the number of $e$-foldings, $\delta N$, is the difference between the curvature perturbations on the initial and final hypersurfaces. We wish to calculate primordial perturbations, hence we pick a final uniform density
hypersurface to be at a fixed time during the standard radiation dominated era, for
example during primordial nucleosynthesis. The initial time is arbitrary provided it is
after the Hubble exit time of all relevant scales. It is often convenient to pick this
time to be shortly after Hubble-exit. We introduce the method for calculating the
primordial curvature perturbation in terms of the $\delta N$ formalism in more detail
below.



Here we present a Feynman diagram type approach to calculating the primordial
non-Gaussianity, where every term in the $n$-point function of $\zeta$ has a diagrammatic
representation. Similar diagrams were introduced in the context of non-linear
perturbation theory of large scale structure \cite{Scoccimarro:1995if} and further
developed in \cite{Crocce:2005xy} (see also \cite{Zaballa:2006pv}). It is possible to
read off the corresponding mathematical term in the $n$-point function for every diagram
and it is possible to draw all the diagrams in a systematic way so that all terms are
included. This approach extends to loop corrections of any required order. Although the
loop corrections are expected to be small in general compared to the tree level terms
there is no proof of this, except for a particular loop correction to the bispectrum in
some specific models of inflation \cite{Zaballa:2006pv}. We will see that in general the
loop corrections depend on a large scale cut off; depending on this cut off the loop
correction may contribute significantly to observable quantities, see for example
\cite{Sloth:2006az,Seery:2007wf}. In fact there is a specific model of inflation where
the one loop term gives the dominant and potentially observable contribution to the
bispectrum \cite{Boubekeur:2005fj}. A diagrammatic method to calculate field fluctuations
during inflation was presented in \cite{Musso:2006pt}.

We first introduce the primordial curvature perturbation. In the next two sections of
this paper we assume the initial field perturbations are Gaussian, which in many cases is
a good approximation to make, for example slow-roll inflation. In
Sec.~\ref{sec:realspacediagrams} the rules for drawing diagrams in real space are given,
along with a discussion of why we want to calculate connected rather than disconnected
diagrams. In Sec.~\ref{sec:fourierspacediagrams} the equivalent rules in Fourier space
are given. An application of the Fourier space rules for the power spectrum including
second order loop corrections is given in Sec.~\ref{sec:powerspectrum}. In
Sec.~\ref{sec:renormalisation} we discuss a way to renormalise the vertices of the
diagrams such that all diagrams with dressed vertices can be absorbed, with a proof given
in the appendix. In Sec.~\ref{sec:nG} the assumption of Gaussian initial fields is
dropped and the complete rules presented and applied to several examples. Finally we
conclude in Sec.~\ref{sec:conclusion}.

\section{The primordial curvature perturbation}

We can write the primordial curvature perturbation, $\zeta$, in terms of derivatives of
$N$ with respect to the fields, $N_A=\partial N/\partial \phi^A$ and the field
perturbation, $\varphi^A$, at the initial time $t_{\rm{ini}}$. By neglecting any
additional dependence on $\dot{\phi}^A$ at Hubble-exit we are assuming the field
perturbations on large scales are overdamped and thus we can neglect the decaying mode.
This is valid during slow roll, but also more generally holds for light fields during
inflation. The assumption that $N$ is a function of field values only is also valid in DBI inflation \cite{Huang:2007hh}.
We use the notation that the fields are $\phi^A=\phi_0^A+\varphi^A$, where the
perturbed field $\varphi^A\equiv\delta\phi^A$ satisfies $\la\varphi^A\ra=0$. The
curvature perturbation is given by
\begin{eqnarray}\label{deltaN}
\zeta&=&\delta N-\la\delta N\ra = N_A\varphi^A+\frac12 N_{AB} \left(\varphi^A\varphi^B-
\la\varphi^A\varphi^B\ra \right)+ \frac{1}{3!}N_{ABC}
\left(\varphi^A\varphi^B\varphi^C-\la\varphi^A\varphi^B\varphi^C\ra\right) + \cdots\,, \\
&&\mathrm{where} \qquad \delta N = N_A\varphi^A+\frac12 N_{AB} \varphi^A\varphi^B+
\frac{1}{3!}N_{ABC} \varphi^A\varphi^B\varphi^C + \cdots\,.
\end{eqnarray}
The dummy index $A$ labels the light scalar fields relevant during inflation, and
summation convention is used throughout this paper. It is more convenient to work with
the $\zeta$ defined above since it satisfies $\la\zeta\ra=0$, even though often
$\zeta=\delta N$ is used in the literature. This equation can hence be used to calculate
the primordial $n$-point function of $\zeta$, although the result will depend on the
$n$-point function of the fields at Hubble exit. This has so far been explicitly
calculated for the 2-, 3- and 4-point functions,
\cite{Maldacena:2002vr,Seery:2005gb,Seery:2006vu}.

The connected 2-, 3- and 4-point functions of the fields are defined by
\begin{eqnarray}\label{fieldcorrelators}
\la \varphi^A_{{\mathbf k_1}}\,\varphi^B_{{\mathbf k_2}}\ra &=& C^{AB}(k)\picube
\sdelta{\bkone+\bktwo} \,, \\  \label{3ptfnfields} \la \varphi^A_{{\mathbf
k_1}}\,\varphi^B_{{\mathbf k_2}}\, \varphi^C_{{\mathbf k_3}}  \ra &=&
B^{ABC}(k_1,k_2,k_3) \picube \sdelta{\bkone+\bktwo+\bkthree}\,,  \\
\label{4ptfnfields} \la \varphi^A_{{\mathbf k_1}}\,\varphi^B_{{\mathbf k_2}}\,
\varphi^C_{{\mathbf k_3}}\, \varphi^D_{{\mathbf k_4}} \ra_c &=&
T^{ABCD}(\bkone,\bktwo,\bkthree,\bkfour) \picube
\sdelta{\bkone+\bktwo+\bkthree+\bkfour}\,.
\end{eqnarray}
Note that only the 4-point function (and higher) depend on the direction of the $\bk$
vectors, the 2- and 3-point functions of the fields just depend on the magnitude of the
vectors, $k_i=|\mathbf{k}_i|$. Homogeneity of the random fields implies that the sum of
the $\bk$ vectors must be zero, while isotropy implies that the $n$-point functions are
invariant under reorientations of this closed configuration.

At lowest order in slow roll different fields are uncorrelated at Hubble exit and all
light fields have the same power spectrum \cite{vanTent:2003mn,byrnes2}, so
\begin{equation}\label{Cfreefields}
C^{AB}(k) = \delta^{AB} P(k) \,,
\end{equation}
where $\delta^{AB}$ is the Kronecker delta-function, and the variance per logarithmic
interval in $k$-space is given by
\begin{equation}\label{P}
{\cal P}(k) = \frac{4\pi k^3}{(2\pi)^3} P(k) = \left(\frac{H_*}{2\pi}\right)^2\,,
\end{equation}
where the Hubble parameter $H$ is evaluated at Hubble-exit, $k=(aH)_*$. At zeroth order
in slow-roll parameters, $\mathcal{P}$ is independent of wavenumber, i.e.~we have a scale
invariant spectrum for the field fluctuations. The bispectrum satisfies
$B^{ABC}=\mathcal{O}(\epsilon^{1/2}P^2)$ \cite{Seery:2005gb}, where $\epsilon$ is a
slow-roll parameter, so it is zero at lowest order in slow roll. However the trispectrum
is not zero at lowest order in slow roll \cite{Seery:2006vu},
$T^{ABCD}=\mathcal{O}(P^3)$, but is suppressed by $\mathcal{O}(P^2)$ compared to the
power spectrum. For a discussion of the slow-roll order of higher order $n$-point
functions of the fields see \cite{Jarnhus:2007ia}.


\section{Diagrams for Gaussian field perturbations}

Initially we will assume that all of the fields have a purely Gaussian distribution at an
initial time, e.g.~shortly after Hubble exit. Then every odd $n$-point function of the
fields is zero and every even $n$-point function of the fields can be reduced to a
product of 2-point functions. This assumption is not required and the extension to
non-Gaussian field perturbations will be presented in Sec.~\ref{sec:nG}. However the
approximation of initially Gaussian field perturbations is frequently made in the
literature and is often a good approximation, for example \cite{Lyth:2005qj} shows that
the 3-point function of the fields adds an unobservably small contribution to the
bispectrum around Hubble exit assuming slow roll inflation with a standard kinetic term
of the scalar fields, while \cite{Seery:2006vu} show that the 4-point function of the
fields adds an unobservably small contribution to the trispectrum around Hubble exit
under the same assumptions.

Working to zeroth order in slow roll implies that (\ref{Cfreefields}) holds and that the
3-point function of the fields is zero. So although working to lowest order in slow roll
is not the same as working with Gaussian field perturbations, in practice the conditions
are related. Hence when assuming that the field perturbations have a Gaussian
distribution at Hubble exit we will also work at zeroth order in slow roll. It is likely
that any observable amount of non-Gaussianity can be calculated to sufficiently high
accuracy when making these two approximations, assuming a standard kinetic term.




\subsection{Real space diagrams}\label{sec:realspacediagrams}

Assuming that the fields are uncorrelated Gaussian variables, with identical
distributions, as discussed above, the two point function of the fields is given by
\bea\label{varphicorrelation} \langle \varphi_{x_i}^{A}\varphi_{x_j}^{B} \rangle =
\left\{
\begin{array}{l} \delta^{A B}G(|x_i-x_j|)\qquad\,\,\,\,\mathrm{for}\,\,\, i\neq j,
\\ \delta^{A B}\langle\varphi^2\rangle
\qquad\qquad\qquad \mathrm{for}\,\,\, i=j.
\end{array}
\right. \eea
All of the higher $n$-point functions of the fields can be written in terms of $G$ and
the variance of the fields $\langle\varphi^2\rangle$. The two point function $G$ only
depends on $|x_i-x_j|$ because we are assuming the background is homogenous and
isotropic.

\indent The diagrammatic rules for the connected $n$-point function,
$\la\zeta_{x_1}\zeta_{x_2}\cdots\zeta_{x_n}\ra_c$, are:

\begin{enumerate}
\item Draw $n$ points representing the $n$ spatial points $x_1,\cdots ,x_n$ and connect them with
$r$ propagators (dashed lines) which attach two of the positions $x_i$ (it can be two
distinct positions or it can have both ends attached to the same position). The diagram
should be connected in order to calculate the connected $n$ point function, see
Sec.~\ref{sec:disconnecteddiagrams}. We require $r\geq n-1$ in order to draw a connected
diagram and diagrams with $r=n-1$ are tree level, while those with $r>n-1$ include loop
corrections.

\item Label each end of each propagator with the field indices $A,B,\ldots C$.

\item Assign a factor $N_{AB\cdots C}$ to each spatial point, $x_i$, where the number of
derivatives of $N$ is the number of propagators attached to that $x_i$.

\item Assign a factor of $\delta^{AB} G(|x_i-x_j|)$ to each propagator, where $AB$ are
the appropriate field indices attached to the propagator and $x_i,x_j$ are the positions
at either end of the propagator. If both ends of the propagator are attached to the same
$x_i$ instead assign a factor $\delta^{AB}\la\varphi^2\ra$.

\item Divide by the appropriate numerical factor. Whenever $l$ propagators attach the
same $x_i$ and $x_j$ at both ends this gives a factor of $l!$. Whenever $l$ propagators
dress an $x_i$ this gives an additional factor of $2^{l}$ (as well as the factor of $l!$
due to the previous rule). A propagator dresses an $x_i$ if both ends of the propagator
are attached to the same $x_i$.

\item Add all permutations of the diagrams which is all of the distinct ways to
relabel the spatial points. The number of permutations depends on the symmetries of the
diagram, a diagram with complete symmetry between all of the spatial points has only one
term, while for a diagram with no symmetries between the spatial points there are $n!$
permutations.



\end{enumerate}

\begin{figure}
\scalebox{0.8}{\includegraphics*{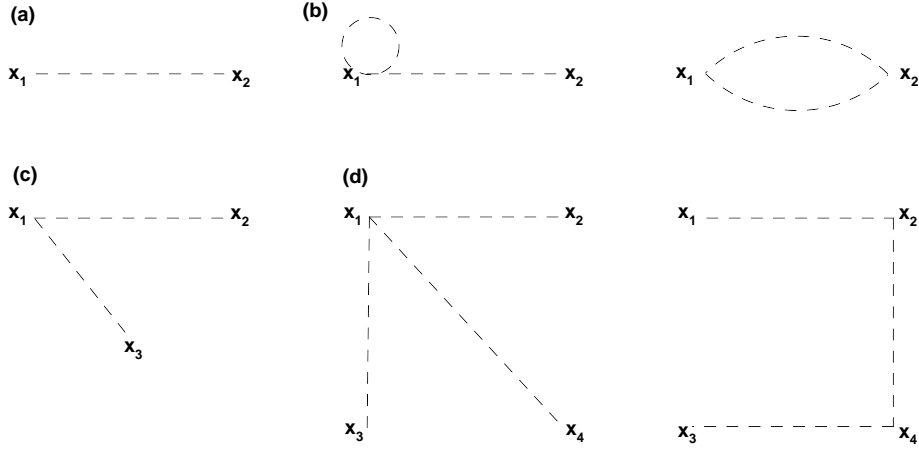}} \caption{The 2-point function at
tree (a) and one loop level (b), and the 3- and 4-point functions at tree level, (c) and
(d) respectively. All figures in this paper were drawn using JaxoDraw
\cite{Binosi:2003yf}.}\label{fig:realspace234pt}
\end{figure}

In Fig.~\ref{fig:realspace234pt} the diagrams for the 2-point function at tree level and
one loop level, plus the tree level terms for the 3- and 4-point functions are shown.
Note that for the 4-point function there are two tree level terms.

The terms corresponding to the diagrams are given below, in the same order as the
diagrams,
\bea \label{2ptfnreal} \la\zeta_{x_1}\zeta_{x_2}\ra &=& \left(N_A N^A +N_A
N^{AB}_B\variancephi\right)G(|x_1-x_2|) +\frac12N_{AB}N^{AB}G(|x_1-x_2|)^2\,,
\\ \la\zeta_{x_1}\zeta_{x_2}\zeta_{x_3}\ra &=& N_A N_B N^{AB} \left( G(|x_1-x_2|)
G(|x_1-x_3|)+2\mathrm{\;perms}\right)\,, \\  \label{4ptfnreal}
\la\zeta_{x_1}\zeta_{x_2}\zeta_{x_3}\zeta_{x_4}\ra_c &=& N_A N_B N_C N^{ABC} \left(
G(|x_1-x_2|) G(|x_1-x_3|) G(|x_1-x_4|) +3\mathrm{\;perms}\right) \nn \\ &&+N_A N_B N^A_C
N^{BC} \left(G(|x_1-x_2|)G(|x_2-x_3|)G(|x_3-x_4|)+11\mathrm{\;perms}\right)\,.  \eea
Although the second diagram for the two point function, Fig.~\ref{fig:realspace234pt},
has a dressed vertex with an associated numerical factor of $2$, there is another
permutation of this diagram with the $x_2$ dressed which gives an equal contribution. The
third diagram has two propagators attaching $x_1$ and $x_2$ and there are no other
permutations of this diagram so there is a numerical factor of $2$, as shown in the third
term of (\ref{2ptfnreal}). The diagrams for the 3- and 4-point functions are all at tree
level so they all have a numerical factor of 1. The first diagram for the 4-point
function has 4 permutations since we have 4 choices of which of the $x_1,\cdots,x_4$ has
3 propagators attached to it as given in the first term of (\ref{4ptfnreal}), the second
diagram of the 4-point function has 12 permutations because there are $4\times 3$ choices
of which 2 of the $x_1,\cdots,x_4$ should have 2 propagators attached to them

\subsubsection{Disconnected diagrams}\label{sec:disconnecteddiagrams}

In general any $n$-point function has a connected and a disconnected part. Throughout
this paper we will only consider the connected $n$-point function because the
disconnected contribution contains no new information over the lower $n$-point functions
that it is split into. There is no disconnected contribution to the 2- or 3-point
function of the fields or zeta, because we assume $\la \varphi^A\ra=\la \zeta\ra=0$. The
disconnected contribution is only non-zero if a subset of the $\bk$ vectors sums to zero.

For example the disconnected part of the primordial 4-point function, $\la
\zeta^A_{{\mathbf k_1}}\,\zeta^B_{{\mathbf k_2}}\,\zeta^C_{{\mathbf k_3}}
\zeta^D_{{\mathbf k_4}}\ra$, is the product of two 2-point functions such as $\la
\zeta^A_{{\mathbf k_1}}\,\zeta^B_{{\mathbf k_2}}\ra \la\zeta^C_{{\mathbf
k_3}}\zeta^D_{{\mathbf k_4}}\ra$ and in total there are three such terms. This is
non-zero even if $\zeta$ is purely Gaussian, but it contains no new information compared
to the primordial 2-point function. This term only contributes if $\bki+\bkj=0$ for some
$i,j=1,\cdots,4$, so provided the sum of any two $\bk$ vectors is not zero any
contribution is from the connected part. The connected part of the primordial 4-point
function is only non-zero if $\zeta$ is not purely Gaussian and it contains information
that is not observable from the primordial 2- or 3-point functions.

\subsection{Fourier space diagrams}\label{sec:fourierspacediagrams}

We present here the rules for drawing the connected diagrams (see
Sec.~\ref{sec:disconnecteddiagrams}) of the $n$-point function of $\zeta$ at $r$-th order
(i.e.~$\mathcal{O}(\P^r)$), for $r\geq n-1$. The tree level terms correspond to $r=n-1$.

\begin{figure}
\scalebox{1}{\includegraphics*{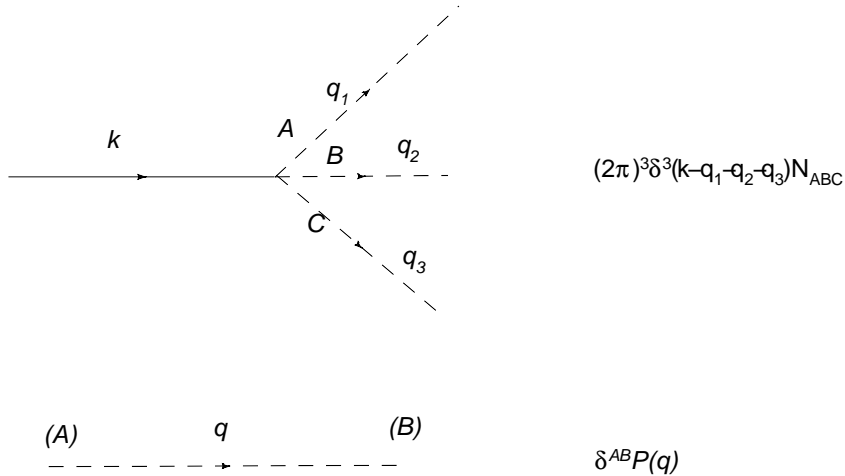}} \caption{The terms that are
associated with every vertex and propagator.}\label{fig:fourierspacerulesG}
\end{figure}

\begin{enumerate}
\item Draw all distinct connected diagrams with $n$-external (solid) lines and $r$ (dashed)
propagators. Every vertex must consist of 1 external line and at least 1 propagator.

\item Label the external legs with incoming momenta $\bki$ for $i=1,\cdots,n$ and label the
propagators with internal momenta $\bqi$ for $i=1,\cdots,r$. Label each end of each
propagator with a field index $A,B,\cdots,C$.

\item  Assign a factor $N_{AB\cdots C}\sdeltapi{k_i-q_1-\cdots-q_p}$ to each vertex. The number of
derivatives of $N$ corresponds to the number of propagators attached to each vertex. We
use the convention that incoming momentum is positive. The $\delta$ function ensures
momentum is conserved at each vertex. See Fig.~\ref{fig:fourierspacerulesG}.

\item Assign a factor $\delta^{AB}P(q)$ to each propagator, where $AB$ are the appropriate field
indices that the propagator is labelled with at either vertex and $\bq$ is the momentum
attached to the propagator. See Fig.~\ref{fig:fourierspacerulesG}.

\item Integrate over the propagator momenta, $\intm q_i$. The first $n-1$ integrations are
trivial because of the $\delta$ functions but any further integrations (in the case of a
diagram with loop corrections) cannot in general be evaluated analytically.

\item Divide by the appropriate numerical factor. Whenever $l$ propagators attach the
same vertices at both ends this gives a factor of $l!$. Whenever $l$ propagators dress a
vertex this gives an additional factor of $2^{l}$ (as well as the factor of $l!$ due to
the previous rule). A propagator dresses a vertex if both ends of the propagator are
attached to the same vertex.

\item Add all permutations of the diagrams which is all of the distinct ways to
relabel the $\bki$ attached to the external lines. The number of permutations depends on
the symmetries of the diagram, a diagram with complete symmetry between all of the
external lines has only one term, while for a diagram with no symmetries between the
external lines there are $n!$ permutations.


\end{enumerate}

\subsubsection{Comments on the diagrams and their rules}\label{sec:comments}

Going between the unlabelled diagrams in Fourier or real space is straightforward, to go
from Fourier space to real space one simply removes the external lines and places an
$x_i$ at each vertex. To go from real space to Fourier space one attaches an external
line in place of every $x_i$.

In Fourier space rule 1 for the diagram is to draw every possible diagram at the
appropriate order. Rule 2 gives the rules for labelling the diagrams. Rules 3--5 give the
associated mathematical expression for each diagram, up to a numerical factor given in
rule 6. A derivation of the numerical factors is given in Sec.~\ref{sec:renormalisation}
and App.~\ref{sec:app:nptfn}. The numerical factor for all tree level diagrams is 1.
Finally rule 7 tells one to include every permutation of the labelling of the $\bki$ on
the external legs that are distinct.

\subsection{Power spectrum}\label{sec:powerspectrum}

\begin{figure}
\scalebox{1}{\includegraphics*{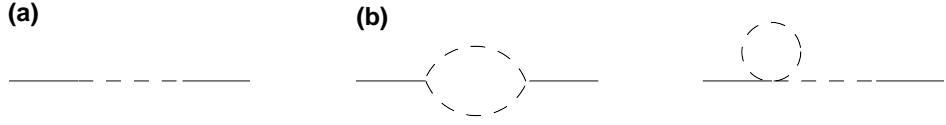}} \caption{The tree level (a) and 1 loop
correction (b) for the power spectrum.}\label{fig:2pt1loopG}
\end{figure}

The primordial power spectrum in Fourier space is defined by
\begin{eqnarray}\label{powerspectrumdefn} \la\zeta_{\bkone}\zeta_{\bktwo}\ra \equiv P_\zeta(k)\picube
\sdelta{\bkone+\bktwo} \,. \end{eqnarray}
The tree level and one loop correction terms to the 2-point function is diagrammatically
given by Fig.~\ref{fig:2pt1loopG}.
After carrying out the trivial integration over the propagator momentum the tree level
term is given by
\begin{eqnarray}\label{2ptfntree}
P_{\zeta}^{\mathrm{tree}}(k)= N_AN^A P(k)\,.
\end{eqnarray}
The one loop terms are given by
\bea\label{2ptfn1loop} P_{\zeta}^{1\mathrm{\;loop}}(k)=\frac{1}{\picube}\int d^3q
\left(\frac12N_{AB}N^{AB}P(q)P(|\bkone-\bq|)+ N_AN^{AB}_BP(k)P(q)\right)\,, \eea
where we have already carried out one integration. Both of the diagrams associated with
these terms have a numerical factor of 2, however there are also 2 permutations of the
second term because it has a non symmetric diagram. Because of momentum conservation,
$\bkone+\bktwo=0$, it follows that $k_1=k_2$.  After enforcing $k\equiv k_1=k_2$ the two
permutations of the second diagram give an equal contribution which cancels the numerical
factor of 2.

\begin{figure}
\scalebox{1}{\includegraphics*{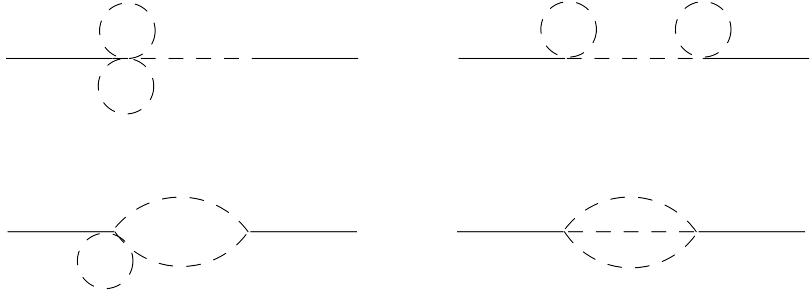}} \caption{The two loop terms for the power
spectrum with a Gaussian initial field.}\label{fig:2pt2loopG}
\end{figure}

Going to 2 loops, terms of order $\P^3$ there are four terms. Diagrammatically the terms
are given in Fig.~\ref{fig:2pt2loopG}
which corresponds to
\bea\label{2ptfn2loop}  P_{\zeta}^{2\mathrm{\;loop}}&=&\frac{1}{(2\pi)^6}\int
d^3q_1d^3q_2 \left(
\frac14N_{ABC}^{AB}N^CP(q_1)P(q_2)P(k)+\frac14N_{AB}^AN^{BC}_CP(k)P(q_1)P(q_2) \right.
\nn \\ && \left. +\frac12N_{AB}N^{ABC}_CP(q_1)P(q_2)P(|\bkone-\bqtwo|)
+\frac16N_{ABC}N^{ABC}P(q_1)P(|\bqtwo-\bqone|)P(|\bkone-\bqtwo|) \right)\,.  \eea

Only the fourth diagram of Fig.~\ref{fig:2pt2loopG} does not have a dressed vertex. In
fact at every level there is only one diagram without dressed vertices, as has already
been shown explicitly at tree and one loop level in Fig.~\ref{fig:2pt1loopG}.

\subsection{Evaluation of the 1 loop diagrams}

In the case where a single field  $\phi$ generates the primordial curvature perturbation,
so $N=N(\phi)$, we have
\bea  P_{\zeta}^{\rm{tree}}(k)=(N')^2 P(k)\,, \eea

where $N'=\partial N/\partial\phi$. In the case of a scale invariant spectrum we can also
evaluate the 1 loop integrals if we apply a large scale cut off $L$. For a discussion of
how to evaluate the integral over the loop momenta and the dependence of this term on the
cut off $L$ see for example \cite{Lyth:2007jh}. We also apply a small scale cut off $k$. The result is
\bea P_{\zeta}^{1\,\,\rm{loop}}(k)= P(k)\P\left((N'')^2+N'''N'\right) \log(kL)\,, \eea
%


%
%
where the variance $\P$ was defined by Eq.~(\ref{P}). In analogy with Eq.~(\ref{P}) we
define $\P_{\zeta}=P_{\zeta}(k)k^3/(2\pi^2)$ which from observations of the CMB is of the
order $10^{-10}$ \cite{Spergel:2006hy}. We hence have
\bea\label{P1loop} \frac{P_{\zeta}^{\rm{1\,\,loop}}(k)}{P_{\zeta}^{\rm{tree}}(k)}
=\P_{\zeta}^{\rm{tree}} \left(\frac{(N'')^2}{(N')^4}+\frac{N'''}{(N')^3}\right)
\log(kL)\,. \eea
We can write the derivatives of $N$ in terms of the (in principle) observable
non-linearity parameters $f_{NL}$ and $g_{NL}$ which in this case can be defined by
\cite{Komatsu:2001rj,Sasaki:2006kq}
\bea
 \label{fandgdefn}
 \zeta=\zeta_1+\frac35f_{NL}\zeta_1^2+\frac{9}{25}g_{NL}\zeta_1^3+\cdots\,,
 \eea
where $\zeta_1$ is Gaussian because it is directly proportional to the initial Gaussian
field perturbation, $\varphi_1$, and the dimensionless non-linearity parameters, $f_{NL}$
and $g_{NL}$, are given by
\bea
 \label{fNL1field}
  f_{NL}&=&\frac56\frac{N''}{(N')^2}\,, \\
 \label{gNL1field}
  g_{NL}&=&\frac{25}{54}\frac{N'''}{(N')^3}\,.
\eea
For the extension of these two formula to the case of multiple field inflation see for
example \cite{byrnes3}. Substituting (\ref{fNL1field}) and (\ref{gNL1field}) into
(\ref{P1loop}) we find
\bea\label{P1loopfg} \frac{P_{\zeta}^{\rm{1 loop}}(k)}{P_{\zeta}^{\rm{tree}}(k)}
=\P_{\zeta}^{\rm{tree}} \left(\frac{36}{25}f_{NL}^2+\frac{54}{25}g_{NL}\right)
\log(kL)\,. \eea
If $\P_{\zeta}^{\rm{tree}}\sim 10^{-10}$ and we take the large scale cut off to be
comparable to the present day Hubble scale then we can take $\log(kL)=O(1)$
\cite{Boubekeur:2005fj,Lyth:2007jh}, and the observational bound on the bispectrum,
$|f_{NL}|\lesssim 100$ \cite{Spergel:2006hy} bounds the first term of the ratio above to
be less than $10^{-6}$. There is not yet any equivalent observational bound on $g_{NL}$
but it would have to be extremely large in order to give a significant contribution to
Eq.~(\ref{P1loopfg}). It therefore seems that the only way for the 1 loop contribution to
be significant compared to the tree level term of the power spectrum is to take an
exponentially large cut off $L$. For a discussion of this possibility see for example
\cite{Sloth:2006az,Seery:2007wf}.

\section{Renormalisation}\label{sec:renormalisation}

There is a way to reduce the number of diagrams significantly, by renormalising the
vertices such that the diagrams with dressed vertices are automatically accounted for. We
do this by renormalising the factors $N_{AB\cdots C}$, that are attached to each vertex.

The derivative of the number of $e$-foldings for the given background $\phi_0$ is
$N_{AB\cdots C}\equiv N_{AB\cdots C}|_{\phi_0}$, and we can relate this to the number of
$e$-foldings at a general point $\mathbf{x}$ by
\bea\label{Ntdefn} \Nt\equiv N(\phi(\x)) =
N|_{\phi_0}+N_A\varphi^A+\frac{1}{2}N_{AB}\varphi^A\varphi^B
+\frac{1}{3!}N_{ABC}\varphi^A\varphi^B\varphi^C+\cdots\,. \eea
The expectation of $\Nt_{AB\cdots C}$ is given by
\bea\label{Nt} \la\Nt_{AB\cdots C}\ra=N_{AB\cdots C}+ \frac{1}{2} N_{AB\cdots
CD}^D\la\varphi^2\ra+\frac18 N_{AB\cdots CDE}^{DE}\la\varphi^2\ra^2+\cdots\,, \eea
where we have used Wick's theorem to decompose the expectation of powers of $\varphi$
according to
\bea\label{wick} \la\varphi^{2n}\ra=(2n-1)!!\la\varphi^2\ra^n, \qquad
\la\varphi^{2n-1}\ra=0\,. \eea
The double factorial is defined by $(2n-1)!!=(2n-1)(2n-3)\cdots 5\cdot3\cdot1$. The
coefficient of the general term in (\ref{Nt}), $\variancephi^n$ is given by
(\ref{Ntdefn}) and (\ref{wick}) and is $(2n-1)!!/(2n)!=1/(2^n n!)$. This explains the
numerical factor associated with every dressed vertex, see rule 6,
Sec.~\ref{sec:fourierspacediagrams}.

In Fourier space the ensemble average of $\Nt_{AB\cdots C}$ is
\bea\label{NtFourier} \la\Nt_{AB\cdots C}\ra=N_{AB\cdots C}+ \frac{1}{2} N_{AB\cdots
CD}^D \frac{1}{(2\pi)^3}\int d^3qP(q) +\frac18 N_{AB\cdots CDE}^{DE} \frac{1}{(2\pi)^6}\int d^3q_1 d^3q_2
P(q_1)P(|\bqtwo-\bqone|) +\cdots\,, \eea
which in real space is given by Eq.~(\ref{Nt}).

\begin{figure}
\scalebox{1}{\includegraphics*{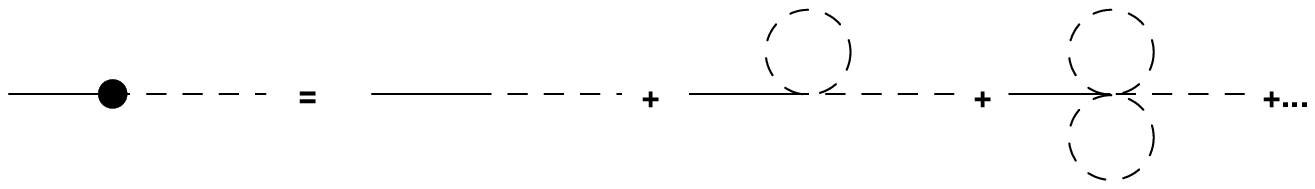}} \caption{Graphical representation of
$\la\Nt_A\ra$.}\label{fig:Ntprimed}
\end{figure}

In Fig.~\ref{fig:Ntprimed} the graphical representation of $\la\Nt_A\ra=N_A+ \frac12
N_{AB}^B\la\varphi^2\ra+\frac18 N_{ABC}^{BC}\la\varphi^2\ra^2+\cdots$ is given up to 2
loops. So far the derivatives of $N$ had been calculated at $\phi_0$, however if we
instead calculate the expectation of the derivative of $N$ at a general point, and attach
this factor $\la\Nt_{AB\cdots C}\ra$ to each vertex (which we call a renormalised vertex)
then all diagrams with dressed vertices are automatically included in the expansion
(\ref{Nt}) and don't need to be drawn. We provide a proof of this in appendix
\ref{sec:app:renN}.

%
%

\subsection{Renormalised real space diagrams}\label{sec:renormalisedrealspacediagrams}

\begin{figure}
\scalebox{1}{\includegraphics*{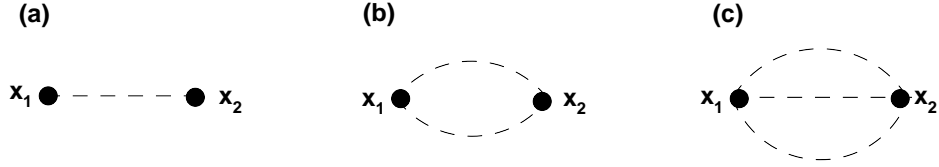}} \caption{The 2-point function at tree,
1 loop and 2 loop levels ((a), (b) and (c) respectively), with renormalised
propagators.}\label{fig:ren2ptfn2loop}
\end{figure}

\indent The new diagrammatic rules for the connected $n$-point function with renormalised
vertices are:

\begin{enumerate}
\item Draw $n$ points representing the $n$ spatial points $x_1,\cdots ,x_n$ and connect them with
$r$ propagators (dashed lines) which attach two distinct positions $x_i$ and $x_j$. The
vertices are drawn with a solid dot to show that they are renormalised, see
Fig.~\ref{fig:ren2ptfn2loop} for examples of such diagrams. The diagram should be
connected in order to calculate the connected $n$ point function, see
Sec.~\ref{sec:disconnecteddiagrams}. We require $r\geq n-1$ in order to draw a connected
diagram and diagrams with $r=n-1$ are tree level, while those with $r>n-1$ include loop
corrections.

\item Label each end of each propagator with the field indices $A,B,\ldots C$.

\item Assign a factor $\la\Nt_{AB\cdots C}\ra$ to each spatial point, $x_i$, where the number of
derivatives of $\Nt$ is the number of propagators attached to that $x_i$.

\item Assign a factor of $\delta^{AB} G(|x_i-x_j|)$ to each propagator, where $AB$ are
the appropriate field indices attached to the propagator and $x_i,x_j$ are the positions
at either end of the propagator.

\item Divide by the appropriate numerical factor. Whenever $l$ propagators attach the
same $x_i$ and $x_j$ at both ends this gives a factor of $l!$.


\item Add all permutations of the diagrams which is all of the distinct ways to
relabel the spatial points. The number of permutations depends on the symmetries of the
diagram, a diagram with complete symmetry between all of the spatial points has only one
term, while for a diagram with no symmetries between the spatial points there are $n!$
permutations.

\end{enumerate}

In Fig.~\ref{fig:ren2ptfn2loop} the diagrams for the 2-point function up to two loop
level is presented.

The terms corresponding to the diagrams are given below, in the same order as the
diagrams,
\bea \la\zeta_{x_1}\zeta_{x_2}\ra &=& \la\Nt_A\ra \la\Nt^A\ra G(|x_1-x_2|)
+\frac12\la\Nt_{AB}\ra \la\Nt^{AB}\ra G(|x_1-x_2|)^2   \nn \\ &&
+\frac{1}{3!}\la\Nt_{ABC}\ra\la\Nt^{ABC}\ra G(|x_1-x_2|)^3 +\cdots\,. \eea
Note that with the renormalised diagrams there is only one diagram at every order in
loops. This result is also derived to all orders in App.~\ref{sec:app:renN}.

\subsection{Renormalised Fourier space diagrams}\label{sec:renormalisedfourierspacediagrams}

Rules for drawing the connected diagrams (see Sec.~\ref{sec:disconnecteddiagrams}) of the
$n$-point function of $\zeta$ at $r$-th order (i.e.~$\mathcal{O}(\P^r)$), for $r\geq
n-1$. The tree level terms correspond to $r=n-1$.

\begin{enumerate}
\item Draw all distinct connected diagrams with $n$ external (solid) lines and $r$ (dashed)
propagators. Every renormalised vertex must consist of 1 external line and at least 1
propagator, and drawn with a solid dot to show that the vertex is renormalised. The
propagators cannot have both ends attached to the same vertex.

\item Label the external legs with incoming momenta $\bki$ for $i=1,\cdots,n$ and label the
propagators with internal momenta $\bqi$ for $i=1,\cdots,r$. Label each end of each
propagator with a field index $A,B,\cdots,C$.

\item  Assign a factor $\la\Nt_{AB\cdots C}\ra\,\sdelta{k_i-q_1-\cdots-q_p}$ to each vertex. The number of
derivatives of $\Nt$ corresponds to the number of propagators attached to each vertex. We
use the convention that incoming momentum is positive. The $\delta$ function ensures
momentum is conserved at each vertex.

\item Assign a factor $\delta^{AB}P(q)$ to each propagator, where $AB$ are the appropriate field
indices that the propagator is labelled with at either vertex and $\bq$ is the momentum
attached to the propagator.

\item Integrate over the propagator momenta, $\intm q_i$. The first $n-1$ integrations are
trivial because of the $\delta$ functions but any further integrations (in the case of a
diagram with loop corrections) cannot in general be evaluated analytically.

\item Divide by the appropriate numerical factor. Whenever $l$ propagators attach the
same vertices at both ends this gives a factor of $l!$.

\item Add all permutations of the diagrams which is all of the distinct ways to
relabel the $\bki$ attached to the external lines. The number of permutations depends on
the symmetries of the diagram, a diagram with complete symmetry between all of the
external lines has only one term, while for a diagram with no symmetries between the
external lines there are $n!$ permutations.

\end{enumerate}

\subsection{Power spectrum}

\begin{figure}
\scalebox{0.9}{\includegraphics*{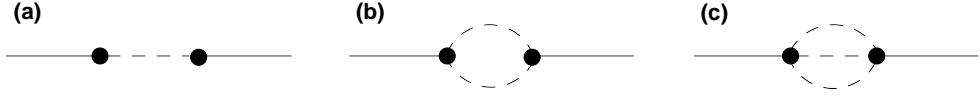}} \caption{The power spectrum at tree, (a),
1, (b), and 2, (c), loop levels with renormalised propagators.}\label{fig:ren2ptfn}
\end{figure}

We again calculate the power spectrum up to 2 loop level, this time with renormalised
vertices. There is only one diagram at every loop level, we present the three lowest
order diagrams in Fig.~\ref{fig:ren2ptfn}, which correspond to
\bea P^{\mathrm{up\;to\;2\;loop}}_{\zeta}&=& \la\Nt_A\ra\la\Nt^A\ra P(k) +\frac{1}{2}
\frac{1}{\picube}\int d^3q \la \Nt_{AB}\ra\la\Nt^{AB}\ra P(q)P(|\bkone-\bq|) \nn \\ && +
\frac{1}{3!} \frac{1}{(2\pi)^6}\int d^3q_1 d^3q_2 \la \Nt_{ABC}\ra\la\Nt^{ABC}\ra
P(q_1)P(|\bqtwo-\bqone|)P(|\bqtwo-\bkone|)\,. \eea

\subsection{Bispectrum}

\begin{figure}
\scalebox{0.8}{\includegraphics*{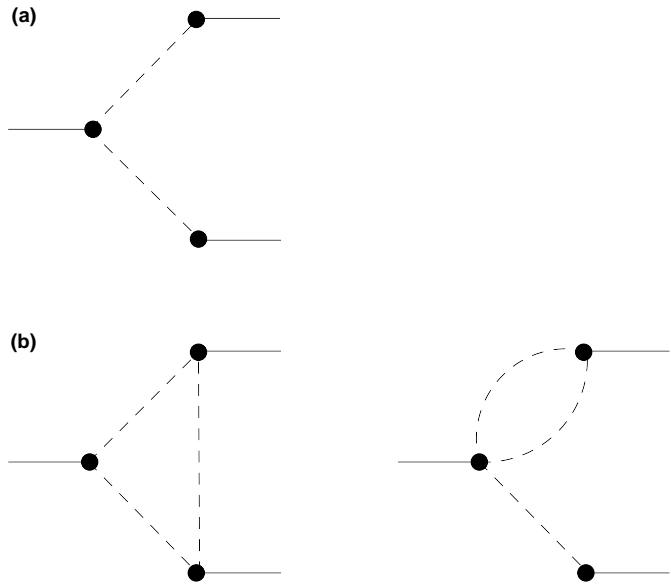}} \caption{The bispectrum at tree,
(a), and 1 loop, (b), levels, with renormalised propagators.}\label{fig:ren3ptfn1loop}
\end{figure}

We calculate the bispectrum including the leading order loop corrections. The bispectrum
is defined by
\bea\label{bispectrum} \langle\zeta_{{\mathbf k_1}}\,\zeta_{{\mathbf k_2}}\,
\zeta_{{\mathbf k_3}}\rangle \equiv B_\zeta( k_1,k_2,k_3) \picube \sdelta{{\mathbf
k_1}+{\mathbf k_2}+{\mathbf k_3}} \,. \eea
At tree level there is one term and at 1 loop level there are two terms, which are shown
diagrammatically by Fig.~\ref{fig:ren3ptfn1loop} and these give the contribution
\begin{eqnarray}\label{ren3ptfn1loop}
B_{\zeta}^{\mathrm{tree}}&=& \la \Nt_{AB}\ra\la\Nt^{A}\ra\la\Nt^{B}\ra
\left(P(k_1)P(k_2)+2\mathrm{\;perms}\right)\,, \\ B_{\zeta}^{1 \mathrm{\;loop}}&=&
\frac{1}{\picube}\int d^3q \bigg( \la\Nt_{AB}\ra \la\Nt^A_C\ra \la\Nt^{BC}\ra
P(q)P(|\bkone-\bq|)P(|\bktwo+\bq|)  \nn
 \\ &&  +\frac12\la\Nt_{ABC}\ra
\la \Nt^A\ra \la\Nt^{BC}\ra \left(P(k_1)P(q)P(|\bktwo-\bq|)+5\mathrm{\;perms}\right)
\bigg) \,.
\end{eqnarray}

\section{Extension to non-Gaussian fields in Fourier space}\label{sec:nG}

If we don't assume that the scalar fields are Gaussian at the initial time (e.g.~Hubble
exit) we have to take account of the higher $n$-point functions of the fields. This is
likely to be important in models with a non-standard kinetic term where the
non-Gaussianity of the fields may be large at Hubble exit \cite{Chen:2006xj}, for example
DBI inflation inspired from brane world models, \cite{DBI,Huang:2007hh}. The connected 2-, 3- and
4-point function of the fields are defined in
(\ref{fieldcorrelators})--(\ref{4ptfnfields}).

Our analysis relies on the $\delta N$-formalism which assumes that $N$ is a
function of the field values $\phi^A$ but not the field derivatives independently as
well. However we can still go beyond leading order in a slow-roll expansion and derive expressions which are valid to arbitrary order in slow roll, for example
the different fields are in general correlated at Hubble exit, at first order in slow
roll, see \cite{vanTent:2003mn,byrnes2}.


\subsubsection{Extended rules for the diagrams with non-Gaussian fields}

The diagrams are more complicated since while the 2-point function is still drawn as a
propagator the higher $n$-point functions are drawn by an $n$-point vertex of dashed
lines.

\begin{figure}
\scalebox{1}{\includegraphics*{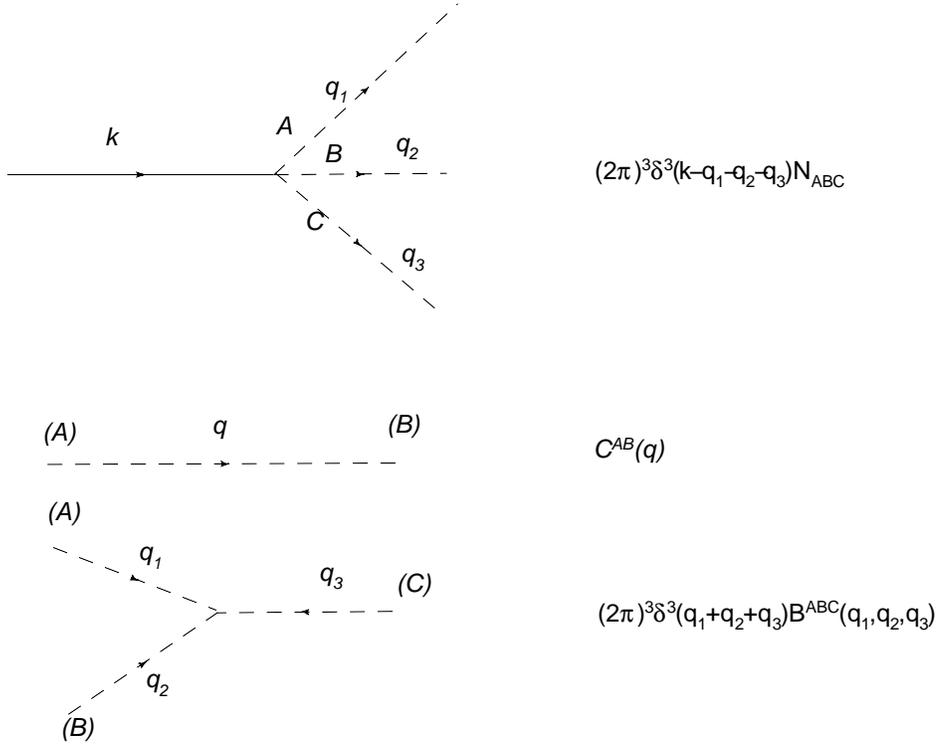}} \caption{The terms that are
associated with every vertex, propagator and higher order correlator.}
\label{fig:fourierspacerulesnG}
\end{figure}

The rules for the diagrams are now:

\begin{enumerate}

\item Draw all distinct connected diagrams with $n$-external (solid) lines and the
appropriate number of internal correlators, which can be propagators (2-point functions)
or higher $n$-point correlators of the fields. The order of the connected $n$-point
function of the field perturbations is $n-1$,
i.e.~$\la\varphi^n\ra_c\sim\mathcal{O}(\P^{n-1})$. There cannot be more than one external
line attached to each vertex. Every dashed line must be attached to an external line on
at least one end, if it is attached to an external line on both ends it is a propagator
(a correlator of order 2).

\item Label the external lines with incoming momenta $\bki$ for $i=1,\cdots,n$ and label
each dashed line with internal momenta $\bqi$ for $i=1,\cdots,r$. Label each dashed line
at a vertex containing an external line with a field index $A,B,\cdots C$.

\item  Assign a factor $N_{AB\cdots C} \sdeltapi{k-q_1-\cdots-q_p}$ to each vertex which
includes an external line, where the number of derivatives of $N$ corresponds to the
number of dashed lines attached to that vertex. The $\delta$ function ensures momentum is
conserved at each vertex. See Fig.~\ref{fig:fourierspacerulesnG}.

\item Assign a factor $C^{AB}(q)$ to each propagator, where $AB$ are the
appropriate field indices that the propagator is labelled with at either vertex and $\bq$
is the momentum attached to the propagator. See Fig.~\ref{fig:fourierspacerulesnG}.

\item Assign a factor $\sdeltapi{q_1+q_2+q_3}B^{ABC}(q_1,q_2,q_3)$ to each vertex of three
dashed lines and no external lines (which corresponds to a 3-point function of the
fields), where $A,B$ and $C$ are the field indices attached to the dashed lines where
they meet a vertex with an external line. See Fig.~\ref{fig:fourierspacerulesnG} again.
Similar rules hold for vertices of dashed lines at higher order.

\item Integrate over all internal momenta, $\intm q_i$. For a tree diagram all of the
integrations are trivial, but in the case of a diagram with loop corrections there will
be some integrals which in general cannot be evaluated analytically.

\item Divide by the appropriate numerical factor. Whenever $l$ propagators or higher
order correlators of the fields attach the same vertices (and with the same number of
lines to each vertex) at both ends this gives a factor of $l!$. If $u$ legs from a higher
order correlator are attached to the same vertex this gives an extra factor of $u!$. It
follows that if $l$ correlators, each of order $p$, all dress the same vertex then this
gives a numerical factor of $l!(p!)^l$. See App.~\ref{sec:app:nG} for a proof of these
rules and see Fig.~\ref{fig:NtnG} for two example of dressed vertices, the associated
terms with the numerical factors are given in (\ref{NtnGeg1}), (\ref{NtnGeg2}).


\item Add all permutations of the diagrams which is all of the distinct ways to
relabel the $\bki$ attached to the external lines. The number of permutations depends on
the symmetries of the diagram, a diagram with complete symmetry between all of the
external lines has only one term, while for a diagram with no symmetries between the
external lines there are $n!$ permutations.

\end{enumerate}

\subsection{Power spectrum}\label{sec:nGpowerspectrum}

We give again the power spectrum defined by (\ref{powerspectrumdefn}), but this time for
non-Gaussian initial fields and valid to all orders in slow roll. There is one extra
diagram at one loop level, and six extra diagrams at two loop level. Because of momentum
conservation, $\bkone+\bktwo=0$, it follows that $k_1=k_2\equiv k$. Therefore even the
non-symmetric diagrams which would have two permutations only correspond to a single term
after enforcing this symmetry.

The tree level term is given by
\begin{eqnarray}
P_{\zeta}^{\mathrm{tree}}(k)=N_AN_BC^{AB}(k)\,,
\end{eqnarray}
notice that at leading order in slow roll using (\ref{Cfreefields}) this reduces to
(\ref{2ptfntree}).

\begin{figure}
\scalebox{1}{\includegraphics*{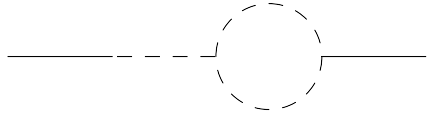}} \caption{The extra one loop term for the
power spectrum with a non-Gaussian field.}\label{fig:2pt1loopnG}
\end{figure}

\begin{figure}
\scalebox{1}{\includegraphics*{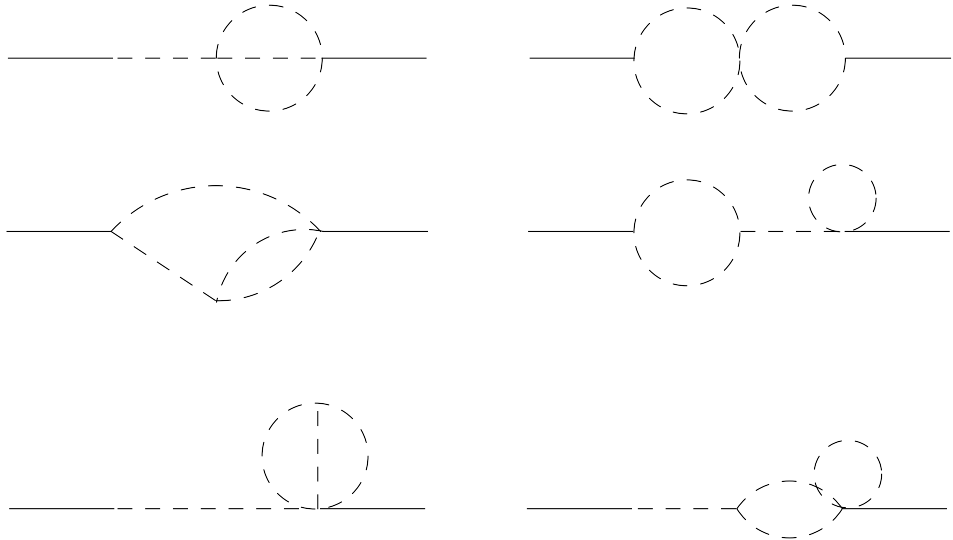}} \caption{The two loop terms for the
power spectrum which involve a 4-point function of the fields.}\label{fig:2pt2loopTBnG}
\end{figure}

The one loop terms are given by
\bea P_{\zeta}^{1\mathrm{\;loop}}(k)&=&\frac{1}{\picube}\int d^3q \bigg(
\frac12N_{AB}N_{CD}C^{AC}(q)C^{BD}(|\bkone-\bq|)+ N_AN_{BCD}C^{AB}(k)C^{CD}(q) \nn \\
&&+N_AN_{BC}B^{ABC}(k,q,|\bkone-\bq|) \bigg)\,, \eea
where the extra one loop term which does not appear for Gaussian fields is shown in
Fig.~\ref{fig:2pt1loopnG}, and the first two terms were shown diagrammatically in
Fig.~\ref{fig:2pt2loopG}. Going to 2 loops, terms of order $\P^3$ there are ten terms,
only the last four of which are non-zero for a Gaussian field. Diagrammatically the terms
involving a 3-point or 4-point function of the fields are given in
Fig.~\ref{fig:2pt2loopTBnG}, which corresponds to the first six terms of
\bea\label{2ptfn2loopnG} \nn P_{\zeta}^{2\mathrm{\;loop}}&=&\frac{1}{(2\pi)^6}\int
d^3q_1d^3q_2 \left(
\frac13N_AN_{BCD}T^{ABCD}(\bqone,\bqtwo-\bqone,\bkone-\bqtwo,\bktwo) \right.  \\
&&+\frac14N_{AB}N_{CD}T^{ABCD}(\bqone,\bkone-\bqone,\bqtwo,\bktwo-\bqtwo)  \nn \\
&&+ N_{AB}N_{CDE}B^{BDE}(|\bkone+\bqone|,|\bqtwo-\bqone|,|\bktwo-\bqtwo|)C^{AC}(q_1)
+\frac12N_{AB}N_{CDE}B^{ABC}(k,q_1,|\bktwo+\bqone|)C^{DE}(q_2)   \nn   \\
&& +\frac13N_AN_{BCDE}B^{CDE}(q_1,|\bqone+\bqtwo|,q_2)C^{AB}(k)
+\frac12N_AN_{BCDE}B^{ABC}(q_1,|\bkone-\bqone|,k)C^{DE}(q_2) \nn   \\  && +
\frac14N_{A}N_{BCDEF}C^{AB}(k)C^{CD}(q_1)C^{EF}(q_2)
+\frac14N_{ABC}N_{DEF}C^{AD}(k)C^{BC}(q_1)C^{EF}(q_2) \nn \\
&&  +\frac12N_{ABCD}N_{EF}C^{AB}(q_1)C^{CE}(|\bkone-\bqtwo|)C^{DF}(q_2) \nn \\
&& \left.
+\frac16N_{ABC}N_{DEF}C^{AD}(q_1)C^{BE}(|\bqtwo-\bqone|)C^{CF}(|\bqtwo-\bkone|)\right)\,.
\eea
The final four terms were already shown diagrammatically in Fig.~\ref{fig:2pt2loopG},
they correspond to the last four terms of (\ref{2ptfn2loopnG}). However the mathematical
expressions for these four diagrams given in (\ref{2ptfn2loop}) are correct at leading
order in slow roll, while the expression above is valid to all orders in slow roll. Note
that 1 of the 1 loop diagrams and 4 of the 2 loop diagrams have dressed vertices, these
can be removed by renormalising the vertices in a similar way to the case with Gaussian
fields, see App.~\ref{sec:app:nG}.

\subsection{Trispectrum}\label{sec:nG4ptfn}

\begin{figure}
\scalebox{1}{\includegraphics{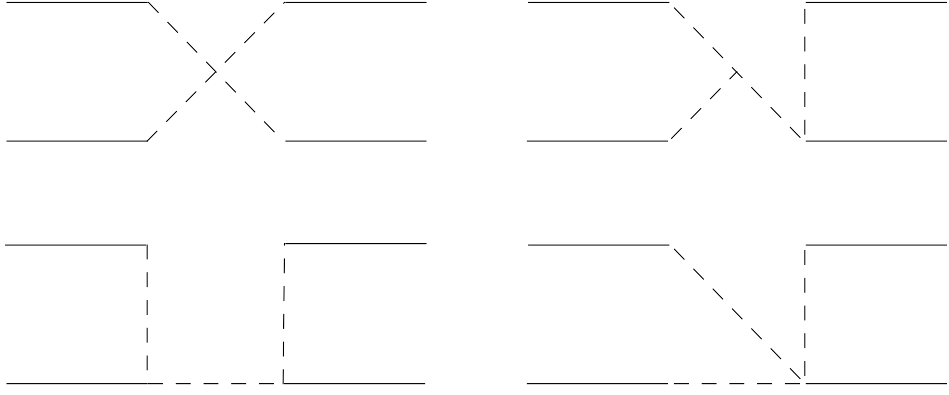}} \caption{Tree level terms of the 4-point
function.}\label{4ptfntreenew}
\end{figure}

The connected part of the primordial trispectrum is defined by
\begin{equation}
\la \zeta_{{\mathbf k_1}}\,\zeta_{{\mathbf k_2}}\, \zeta_{{\mathbf k_3}}\,
\zeta_{{\mathbf k_4}} \ra_c \equiv T_\zeta({\mathbf k_1},{\mathbf k_2},{\mathbf k_3},
{\mathbf k_4}) \picube \sdelta{{\mathbf k_1}+{\mathbf k_2}+{\mathbf k_3} +{\mathbf
k_4}}\,.
\end{equation}
Diagrammatically all of the tree level terms for the 4-point function are given by
Fig.~\ref{4ptfntreenew}.
Writing down these terms in order we get
\bea\label{trispectrumfull} T_\zeta^{\mathrm{tree}}(\bkone,\bktwo,\bkthree,\bkfour)
&=&N_AN_BN_CN_D T^{ABCD}(\bkone,\bktwo,\bkthree,\bkfour) \nn \\ && + N_{AB}N_C N_D N_E
\left( C^{AC}(k_1)B^{BDE}(|\bkone+\bktwo|,k_3,k_4)+ (11\,\, \rm{perms})\right) \nn
\\ && + N_{AB}N_{CD}N_E N_F \left(C^{AC}(|\bkone+\bkthree|)C^{BD}(k_3)C^{DF}(k_4)
+(11\,\, \rm{perms})\right)  \nn \\ && +N_{ABC}N_D N_E N_F
\left(C^{AD}(k_2)C^{BE}(k_3)C^{CF}(k_4)+(3\,\,\rm{perms})\right)\,. \eea
This result was first derived in \cite{Seery:2006js,byrnes3} without using diagrams (see
also \cite{Boubekeur:2005fj}). Note that all of the numerical coefficients are one,
because all of the terms are tree level, and that only the final two terms are non-zero
if the field fluctuations are Gaussian. The real space diagrams for these two terms were
given in Fig.~\ref{fig:realspace234pt}.

\section{conclusion}\label{sec:conclusion}

We have presented a diagrammatic method for calculating any $n$-point function of the
primordial curvature perturbation, $\zeta$, at tree-level or any required loop level.
Rules are given for first drawing diagrams at the required order, and then for writing
down the corresponding terms in the $n$-point function, in either real or Fourier space.
For example, we have drawn diagrams corresponding to all the 1- and 2-loop terms to the
power spectrum and have given the corresponding corrections to the power spectrum.

Our method is based on using the $\delta N$-formalism which identifies the local
curvature perturbation with the perturbation of the integrated expansion from an initial
hypersurface.
The expansion on large scales (typically larger than the Hubble scale) can be determined
as a (non-linear) function of the initial local values of scalar fields during inflation
using the homogeneous (FRW) equations of motion in the separate universe approach.
In practice we are working with fields that are smoothed on scales
$k>k_{\rm{max}}=(aH)_*$. In particular we neglect any ultraviolet divergence of the
fields on small scales.



We show that it is possible to renormalise vertices by absorbing an infinite sum of
terms, corresponding to loop corrections to vertices. These are automatically accounted
for if we work in terms of the average local expansion $\langle N(\phi({\bf x}))\rangle$
and its derivatives, defined in Eq.~(\ref{Nt}) for Gaussian fields, instead of the
background expansion, $N(\phi_0)$.
This removes terms such as $\langle\varphi^2\rangle$ which is divergent for
scale-invariant spectra, leaving all $n$-point functions finite in real space so long as
$\langle N(\phi({\bf x}))\rangle$ and its derivatives are finite.
However divergent terms remain in Fourier space due to integrals over loop momenta
between vertices which may diverge in the infrared.
A discussion of the effect of the large scale cut off on observables is given in
\cite{Boubekeur:2005fj,Lyth:2006gd}.\footnote{Two very recent papers which discuss this
in depth are \cite{Lyth:2007jh} by Lyth who estimates bounds on the size of the loop
corrections to the 2-, 3- and 4-point functions and \cite{Seery:2007we,Seery:2007wf} by
Seery who more explicitly calculates the size of the one-loop contribution to the power
spectrum. He finds that any large loop correction will come from the loop correction to
the curvature perturbation on super Hubble scales rather then the quantum mechanical loop
corrections on sub Hubble scales.}

Results  for the primordial $n$-point function are given in terms of derivatives of $N$
with respect to the fields at Hubble exit and the $n$-point functions of the fields
evaluated at Hubble exit.

In single-field models of inflation the curvature perturbation is constant on super
Hubble scales, so it is possible to calculate the derivatives of $N$ explicitly at Hubble
exit during inflation \cite{byrnes3} without needing to know the details of the end of
inflation or reheating, etc. In this case both the bispectrum and trispectrum have been
shown to be small, regardless of whether the fields are assumed to be Gaussian or not,
but assuming the field has a standard kinetic term in the action. During inflation the
derivatives of $N$ can be calculated explicitly even in multiple field models for a
separable potential, this been done to second
\cite{Vernizzi:2006ve,Battefeld:2006sz,Choi:2007su} and third order \cite{Seery:2006js}.
Several specific multiple field models have been studied in greater detail,
e.g.~$\mathcal N$-flation \cite{Kim:2006te,Battefeld:2007en} and a model of multiple
field inflation in which $N$ can be calculated exactly, \cite{Sasaki:2007ay}. For all
examples considered so far in the slow roll regime the non-Gaussianity has proved to be
small (see also \cite{Alabidi:2005qi,Rigopoulos:2005us,Yokoyama:2007uu}), however there
are many models which explicitly generate a large non-Gaussianity after inflation.
Examples include the curvaton scenario \cite{curvaton} and modulated reheating
\cite{modulatedreheating}, also preheating, for example with hybrid inflation
\cite{Barnaby:2006km}. In general a numerical calculation of the derivatives of $N$ is
required, the advantage of the $\delta N$ formalism is that we only need to solve the
homogenous background equations.

If one assumes the fields are Gaussian then only the 2-point function of the fields is
required, but in general one requires the $(r-1)$-point function of the fields when
working to $r$'th order for any $n$-point function of $\zeta$. So far the 3- and 4-point
functions have been calculated, both at leading order in slow roll and ignoring any
quantum mechanical loop corrections. It is possible but not expected that these
corrections may be important \cite{Sloth:2006az,Weinberg:2006ac}. In the most commonly
considered case of slow-roll inflation, the field is extremely close to Gaussian at
Hubble exit so it is a good approximation to assume that the field is Gaussian and that
any observable non-Gaussianity is generated after Hubble exit.

Methods to generate a large non-Gaussianity during inflation include having a break in
the potential, \cite{Chen:2006xj}, or to have a non-standard kinetic term so that the
non-Gaussianity of the fields is large at Hubble exit \cite{Chen:2006nt}, for example DBI
inflation inspired from brane world models, \cite{DBI}.

\acknowledgments

The authors are grateful to David Lyth and Robert Crittenden for comments. KK is
supported by PPARC (STFC). MS is supported in part by JSPS Grants-in-Aid for Scientific
Research (A) No.~18204024 and (B) No.~1734007. The authors acknowledge support by the
JSPS Japan-UK collaboration program.

\appendix

\section{Renormalising derivatives of $N$}\label{sec:app:renN}

Here we provide a proof that renormalising vertices removes the diagrams with dressed
vertices. In real space this is equivalent to removing terms with $\variancephi$. We
first provide a proof for the power spectrum where the notation is simpler. We then
provide a proof for the general $n$-point function, initially assuming that the fields
are Gaussian at Hubble exit, Sec.~\ref{sec:app:nptfn}, and finally extend this proof to
allow for arbitrary field correlations at Hubble exit, Sec.~\ref{sec:app:nG}.

The gauge invariant curvature perturbation is defined by (\ref{deltaN}) as
\bea \zeta=\delta N-\la\delta N\ra = \sum_{p=1}^{\infty}\frac{1}{p!}
N_{\underbrace{AB\cdots C}_{p}} \left(\varphi^A\varphi^B\cdots\varphi^C-
\la\varphi^A\varphi^B\cdots\varphi^C\ra\right)\,, \eea
where the subscript $p$ means that there are $p$ field indices above the brace. From the
definition of $\zetab$, the 2-point function is given by
\bea\label{app:2ptfn} \la\zetab_x\zetab_y\ra =
\sum_{p,q=1}^{\infty}\frac{1}{p!q!}N_{\underbrace{AB\cdots C}_{p}}N_{\underbrace{DE\cdots
F}_{q}} \la \left( \varphi_x^A\varphi_x^B\cdots\varphi_x^C-\la
\varphi_x^A\varphi_x^B\cdots\varphi_x^C \ra \right)
\left(\varphi_y^D\varphi_y^E\cdots\varphi_y^F- \la \varphi_y^D\cdots\varphi_y^F \ra
\right) \ra\,. \eea
Because the fields are Gaussian every $n$-point function of the fields can be split into
products of 2-point functions. We can choose the pairs to be connected pairs, i.e.~each
$\varphi$ is evaluated at a different point or disconnected pairs which are
$\variancephi$. We consider every possible way of splitting the term $\la
\varphi_x^A\varphi_x^B\cdots\varphi_x^C\varphi_y^D\varphi_y^E\cdots\varphi_y^F\ra$ into
pairs:

-- Only disconnected pairs: this term vanishes from (\ref{app:2ptfn}) because we are
working with $\zeta$ which satisfies $\la\zeta\ra=0$.

-- 1 connected pair: there are $pq$ different ways to select this term. This gives
\bea &&\sum_{p,q}\frac{pq}{p!q!} N_{\underbrace{AB\cdots C}_{p}} N_{\underbrace{DE\cdots
F}_{q}} \delta^{AD} G(|x-y|) \la \varphi_x^B\cdots\varphi_x^C \ra \la
\varphi_y^E\cdots\varphi_y^F \ra=\la\Nt_A\ra\la\Nt^A\ra G(|x-y|)  \,, \\
&&\mathrm{where}\qquad \la\Nt_A\ra=
\sum_{p=1}^{\infty}\frac{1}{(p-1)!}N_{A\underbrace{B\cdots
C}_{p-1}}\la\varphi^B\cdots\varphi^C\ra\,.   \eea

-- 2 connected pairs: there are $p(p-1)q(q-1)/2!$ ways to choose these two connected
pairs. The factor of $2!$ comes because the order in which we pick the pairs does not
matter. This gives
\bea && \frac{1}{2!}\sum_{p,q}\frac{1}{(p-2)!(q-2)!} N_{AB\underbrace{\cdots C}_{p-2}}
N_{DE\underbrace{\cdots F}_{q-2}} \delta^{AD}\delta^{BE} G^2(|x-y|) \la \cdots\varphi_x^C
\ra \la \cdots\varphi_y^F \ra \nn  \\  &=& \frac{1}{2!} \la\Nt_{AB}\ra\la\Nt^{AB}\ra
G^2(|x-y|) \,.  \eea
Continuing this for every possible number of connected pairs, we find that
(\ref{app:2ptfn}) is equivalent to
\bea \la\zetab_x\zetab_y\ra = \sum_{r=1}^{\infty}\frac{1}{r!}\la\Nt_{\underbrace{AB\cdots
C}_{r}}\ra \la\Nt^{AB\cdots C}\ra G^{\,r}(|x-y|)\,. \eea

\subsection{General $n$-point function}\label{sec:app:nptfn}

From the definition of $\zetab$ the $n$-point function is given by
\bea\label{app:nptfn} \la\zetab_x\zetab_y\cdots\zetab_z\ra &=&
\sum_{\underbrace{p,q,\ldots,r}_{n}} \frac{1}{p!q!\cdots r!}N_{\underbrace{AB\cdots
C}_{p}}N_{\underbrace{DE\cdots F}_{q}} \cdots N_{\underbrace{GH\cdots I}_{r}} \nn  \\ &&
\times \la \left(\varphi_x^A\varphi_x^B\cdots\varphi_x^C-
\la\varphi_x^A\varphi_x^B\cdots\varphi_x^C\ra\right) \cdots
\left(\varphi_z^G\varphi_z^H\cdots\varphi_z^I-
\la\varphi_z^G\varphi_z^H\cdots\varphi_z^I\ra \right) \ra \,. \eea

Similar to the calculation of the 2-point function, we have to split the term $\la
\varphi_x^A\varphi_x^B\cdots\varphi_x^C\cdots\varphi_z^G\varphi_z^H\cdots\varphi_z^I\ra$
into pairs in every possible way. In the case that some of the $x,y,\cdots z$ do not
appear in a connected pair there is no contribution to the $n$-point function, for
example if $\varphi_z$ terms do not appear in any connected terms then the last bracket
of (\ref{app:nptfn}) is $\left(\la\varphi_z^G\varphi_z^H\cdots\varphi_z^I\ra-
\la\varphi_z^G\varphi_z^H\cdots\varphi_z^I\ra \right)=0$. We are left with
\bea\label{app:reducednptfn} \la\zetab_x\zetab_y\cdots\zetab_z\ra &=&
\sum_{\underbrace{p,q,\ldots r}_{n}} \frac{1}{p!q!\cdots r!}N_{\underbrace{AB\cdots
C}_{p}}N_{\underbrace{DE\cdots F}_{q}} \cdots N_{\underbrace{GH\cdots I}_{r}} \nn  \\ &&
\times \la \varphi_x^A\varphi_x^B\cdots\varphi_x^C \cdots
\varphi_z^G\varphi_z^H\cdots\varphi_z^I\ra_{cc} \,, \eea
where the subscript $cc$ to the angular brackets means that they must be decomposed into
pairs such that every $x,y,\ldots ,z$ appears in at least one connected pair. We
introduce the notation that $l_{xy}\geq 0$ is the number of connected pairs of $x$ and
$y$ and similarly for every other possible connected pair. Furthermore $n_x$ is the
number of $\varphi_x$ that form part of a connected pair, so $n_x=l_{xy}+\cdots+l_{xz}$.
Note that we require $1\leq n_x \leq p$.

We consider a term in (\ref{app:reducednptfn}) from an arbitrary choice of
$l_{xy},\cdots,l_{xz}$. The number of ways to choose the $l_{xy}$ connected pairs of
$x,y$ correlators is given by $p(p-1)\cdots(p-l_{xy}+1)q\cdots(q-l_{xy}+1)/l_{xy}!$, then
the number of ways to choose the $l_{xz}$ connected correlators linking $x$ and $z$ is
given by $(p-l_{xy})(p-l_{xy}-1)\cdots(p-l_{xy}-l_{xz}+1)\cdots
r(r-1)\cdots(r-l_{xz}+1)/l_{xz}!$. Continuing this counting for all the pairs, we find
the number of ways to split into the required number of connected pairs is
\bea \frac{p(p-1)\cdots(p-n_x+1)\cdots r(r-1)\cdots (r-n_z+1)}{l_{xy}!l_{xz}!\cdots
l_{yz}!}\,. \eea
This choice of $l_{xy},\cdots ,l_{yz}$ therefore gives the following term to
(\ref{app:nptfn}),
\bea && \frac{1}{l_{xy}!\cdots l_{yz}!}
\sum_{p,q,\ldots,r}\frac{1}{(p-n_x)!\cdots(r-n_z)!} N_{AB\cdots C}N_{DE\cdots F} \cdots
N_{GH\cdots I} \delta^{AD}\delta^{BE}\cdots \nn \\   && \times G^{l_{xy}}(|x-y|)\cdots
G^{l_{yz}}(|y-z|)\la \underbrace{\cdots \varphi_x^C}_{p-n_x}\ra
\cdots\la\underbrace{\cdots\varphi_z^I}_{r-n_z}\ra \\ &=& \frac{1}{l_{xy}!\cdots l_{yz}!}
\la\Nt_{\underbrace{AB\cdots C}_{n_x}}\ra\cdots \la\Nt_{\underbrace{GH\cdots I}_{n_z}}\ra
\delta^{AD}\delta^{BE}\cdots G^{l_{xy}}(|x-y|)\cdots G^{l_{yz}}(|y-z|)\,, \eea
where there are $l_{xy}$ contraction of field indices of the first two factors of
$\la\Nt_{AB\cdots C}\ra$ etc. Repeating this for all allowed choices of
$l_{xy},\cdots,l_{yz}$ we get
\bea \la\zetab_x\zetab_y\cdots\zetab_z\ra = \sum_{l_{xy},\cdots,l_{yz}}
\frac{1}{l_{xy}!\cdots l_{yz}!} \la\Nt_{\underbrace{AB\cdots C}_{n_x}}\ra\cdots
\la\Nt_{\underbrace{GH\cdots I}_{n_z}}\ra \delta^{AD}\delta^{BE}\cdots
G^{l_{xy}}(|x-y|)\cdots G^{l_{yz}}(|y-z|)\,. \eea
Diagrammatically a term $G^{l_{xy}}(|x-y|)$ corresponds to $l_{xy}$ propagators which all
have both ends attached to the same spatial points $x$ and $y$. This explains the
numerical factor of $l_{xy}!$, as given by rule 5 in real space (or equivalently rule 6
in Fourier space).

\subsection{Extension to non-Gaussian fields}\label{sec:app:nG}

Here we extend the proof that renormalising vertices eliminates the diagrams with dressed
vertices also holds for non-Gaussian fields. By doing this we also derive the numerical
factors that are associated with every diagram. The proof is more complex than before
because we must consider every possible correlation of an arbitrary number of fields,
rather than reducing every $n$-point correlation to products of 2-point functions. We now
have
\bea\label{app:generalisedNt} \la\Nt_{AB\cdots C}\ra &=& N_{AB\cdots C}+ \frac{1}{2}
N_{AB\cdots CDE}\la\varphi^D\varphi^E\ra +\frac{1}{3!}N_{AB\cdots CDEF}
\la\varphi^D\varphi^E\varphi^F\ra   \nn \\ &&  +\frac{1}{4!} N_{AB\cdots CDEFG}
\la\varphi^D\varphi^E\varphi^F\varphi^G\ra+\cdots\,. \eea
The correlators here includes the connected and disconnected parts.

Again the $n$-point function of $\zetab$ is given by (\ref{app:nptfn}) and we can reduce
it to (\ref{app:reducednptfn}) in a similar way to the case with Gaussian fields. However
we now have to decompose the term in angle brackets into connected correlators of
arbitrary order in every possible way. We extend the notation that $l_{xy}$ is the number
of terms like $\la\varphi_x^A\varphi_y^B\ra$ to more general correlators like $l_{xxzz}$
is the number of terms like $\la\varphi_x^A\varphi_x^B\varphi_z^C\varphi_z^D\ra_c$. Every
$l$ must have at least 2 distinct $x_i$ in the subscript. We define $n_x$ as the number
of $x$'s contained in the correlators and again we require that $1\leq n_x\leq p$ and
similarly for $n_y$, etc. We consider the term in (\ref{app:reducednptfn}) coming from a
particular choice of $l_{xy},\cdots,l_{x\cdots x\cdots z}$, which we denote by
\bea  P^{A\cdots BC\cdots EF\cdots G\cdots H} &=&
\underbrace{\la\varphi_x^A\varphi_y^E\ra\cdots\la\varphi_x^B\varphi_y^F\ra}_{l_{xy}}
\cdots \underbrace{\la\varphi_x^C\cdots\varphi_x^D\cdots\varphi_z^G\ra_c
\cdots}_{l_{x\cdots x\cdots z}} \cdots  \nn \\  &=& \prod
\la\underbrace{\varphi_x\cdots\varphi_x}_{p}\cdots\underbrace{\varphi_z\varphi\varphi_z}_{w}
\ra_c^{l_{x\cdots x\cdots z\cdots z}} \eea
and the product on the last line is over all chosen correlators of the fields and we have
also dropped the field indices on this line. We can repeat the calculation for the number
of ways to split the total correlator of (\ref{app:reducednptfn}) into the appropriate
connected pairs. The only new point is that whenever an $x_i$ appears more then once in
the same $l_{x\cdots x\cdots z}$ the order in which we pick those $x$'s does not matter,
so we have to divide by the factorial of the number of $x$'s raised to the power
$l_{x\cdots x\cdots z}$. In total we get
\bea \frac{p(p-1)\cdots(p-n_x+1)\cdots r(r-1)\cdots (r-n_z+1)}{\prod l_{x\cdots x\cdots
z}!(u!\cdots w!)^{l_{x\cdots x\cdots z}}}\,, \eea
where the product in the denominator is over all $l$'s and the number of $x_i$'s repeated
in each individual $l$.

Therefore this choice of $l$'s gives the following contribution to (\ref{app:nptfn}),
\bea  &&  \frac{1}{\prod l_{x\cdots x\cdots z}!(u!\cdots w!)^{l_{x\cdots x\cdots z}}}
\sum_{p,q,\cdots,r}\frac{1}{(p-n_x)!\cdots(r-n_z)!} N_{AB\cdots C}N_{DE\cdots F} \cdots
N_{GH\cdots I} \nn  \\   &&  \times P^{AB\cdots GH} \la \underbrace{\cdots
\varphi_x^C}_{p-n_x}\ra \cdots\la\underbrace{\cdots\varphi_z^I}_{r-n_z}\ra  \\ &=&
\frac{1}{\prod l_{x\cdots x\cdots z}!(u!\cdots w!)^{l_{x\cdots x\cdots z}}}
\la\Nt_{\underbrace{AB\cdots C}_{n_x}}\ra\cdots \la\Nt_{\underbrace{GH\cdots I}_{n_z}}\ra
P^{AB\cdots GH} \,. \eea
Repeating this for all allowed choices of the $l$'s this gives
\bea\label{app:nptfnnG} \la\zetab_x\zetab_y\cdots\zetab_z\ra =
\sum_{l_{xy},\cdots,l_{x\cdots x \cdots z}} \frac{1}{\prod l_{x\cdots x\cdots
z}!(u!\cdots w!)^{l_{x\cdots x\cdots z}}} \la\Nt_{\underbrace{AB\cdots C}_{n_x}}\ra\cdots
\la\Nt_{\underbrace{GH\cdots I}_{n_z}}\ra P^{AB\cdots I} \,. \eea

\begin{figure}
\scalebox{1}{\includegraphics{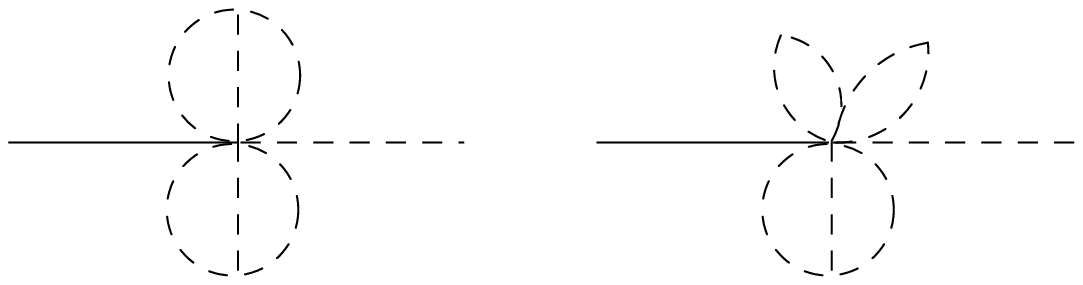}} \caption{Two terms of the renormalised vertex
$\la\Nt_A\ra$.}\label{fig:NtnG}
\end{figure}

Eq.~(\ref{app:nptfnnG}) explains the numerical factor of rule 7 in Sec.~\ref{sec:nG},
except for the case of dressed vertices. The numerical factor for dressed vertices
requires a calculation of the number of ways to reduce the $n$-point function in
(\ref{app:generalisedNt}) into connected pairs of the required order. If $l$ correlators
of order $n$ each dress the same vertex then we need to know the number of ways to split
the $ln$-point function into $l$ lots of connected $n$-point functions. The result is
\bea \frac{1}{l!} \,^{ln}C_n\,^{n(l-1)}C_n\cdots \,^{2n}C_n = \frac{(ln)!}{l!(n!)^l}\,,
\eea
where the denominator of $l!$ arises because the order in which we pick the $n$-point
functions does not matter. The factor of $(ln)!$ is cancelled by the numerical factor in
(\ref{app:generalisedNt}). The generalisation to more complicated dressed vertices is
straightforward to calculate, for example the number of ways to split an
$(n_1l_1+n_2l_2)$-point function into $l_1$ lots of connected $n_1$-point functions and
$l_2$ lots of connected $n_2$-point functions is
\bea \frac{1}{l_1!l_2!}\frac{(l_1n_1+l_2n_2)!}{(n_1!)^{l_1}(n_2!)^{l_2}}\,. \eea
In Fig.~\ref{fig:NtnG} we give two examples of these rules, both of which are 4 loop
terms of the dressed vertex of $\Nt_A$, the first diagram has $l=2$ and $n=3$ while the
second diagram has $l_1=2,n_1=2,l_2=1$ and $n_2=3$. The associated mathematical
expressions are
\bea && \label{NtnGeg1}
\frac{1}{2!3!^2}N_{ABCDEFG}\la\varphi^B\varphi^C\varphi^D\ra\la\varphi^E\varphi^F\varphi^G\ra\,,
\\ && \frac{1}{2!2!^23!}N_{ABCDEFGH}\la\varphi^B\varphi^C\ra\la\varphi^D\varphi^E\ra
\la\varphi^F\varphi^G\varphi^H\ra\,. \label{NtnGeg2} \eea

\end{document}